\documentclass[twocolumn,prc,aps,nofootinbib]{revtex4}

\usepackage[latin9]{inputenc}
\usepackage{color}
\usepackage{amsmath}
\usepackage{amsthm}
\usepackage{amssymb}
\usepackage{graphicx}

\usepackage{ulem}

\makeatletter

\usepackage{epsfig}
\usepackage{amsthm}
\usepackage{tikz}
\usetikzlibrary{quantikz}

\usepackage{bm}

\usepackage{color}

\makeatother

\begin{document}

\title{Restoring broken symmetries using oracles} 

\author{Edgar Andres Ruiz Guzman }
\email{ruiz-guzman@ijclab.in2p3.fr}
\affiliation{Universit\'e Paris-Saclay, CNRS/IN2P3, IJCLab, 91405 Orsay, France}

\author{Denis Lacroix }
\email{denis.lacroix@ijclab.in2p3.fr}
\affiliation{Universit\'e Paris-Saclay, CNRS/IN2P3, IJCLab, 91405 Orsay, France}

\date{\today}

\begin{abstract}
We present a new method to perform variation after projection in many-body systems on 
quantum computers that does not require performing explicit projection. 
The technique employs the notion of ``oracle'', generally used in quantum search algorithms. 
We show how to construct the oracle and the projector associated with a symmetry operator. 
The procedure is illustrated for the parity, particle number, and total spin symmetries. 
The oracle is used to restore symmetry by indirect measurements using a single ancillary qubit. 
An Illustration of the technique is made to obtain the approximate ground state energy for the pairing model Hamiltonian. 
\end{abstract}

\maketitle 

\section{Introduction}

Quantum computers promise to speed up the computation of some selected problems that are hard to solve on classical computers \cite{Nie02,Hid19}. 
One of the opportunities offered by quantum technologies is the simulation of many-body quantum systems with a large number of particles,
where the exponential scaling of their Hilbert space prevents their ab-initio description in classical computers as the number of degrees of freedom increases. 
Assuming we can describe the physical system accurately in the quantum computer, multiple methods have been designed to estimate the ground state energy of a Hamiltonian, e.g., \cite{Per14,McC16,Fan19,Cao19,McA20,Bau20}.
We are now in the Noisy intermediate-scale quantum  (NISQ) computers  era \cite{Pre18,Bha22}; in this period, the algorithms should be tailor-made to handle a limited number of gates and qubits and the presence of noise.
The variational quantum eigensolver (VQE) \cite{Per14,McC16} is one of the currently used best candidates to fill the requirements listed above
due (i) to its comparatively short coherence time and (ii) to the possibility of customizing the ans\"atz to the physical problem at hand.

Because of the inherent noise of current quantum processors, the symmetry that a wave function should respect when solving a physical problem will most likely be broken accidentally. 
A possible way to control the errors is eventually to enforce the symmetry with specific algorithms \cite{Got97,Saw16,Bon18,Sag19,Tra21,Koc21,Hug21}. 
In some cases, like when a system encounters a spontaneous symmetry breaking, it can also be helpful to break some symmetries on purpose \cite{Rin80,Bla86,Ben03,Rob18,She19}. 
In both cases, wanted or unwanted symmetry-breaking (SB), specific methods should be designed to restore the symmetry (SR) in quantum computations. 
In the many-body context, this avenue has been recently explored, requiring \cite{Lac20} or not  \cite{Kha21} the explicit construction of the symmetry projected wave function. 
The method proposed in \cite{Lac20} applies to any symmetries, including spin projection problems \cite{Siw21}. The symmetry projection  
 was used in addition to classical optimization post-processing calculations in Refs.  \cite{Rui21,Rui22}
 to obtain ground state and excited states in many-body systems.  
 In particular, in \cite{Rui21}, the equivalent to the Variation-After-Projection, called Q-VAP, has been applied to superfluid systems. 
 This method is based on the Quantum-Phase-Estimation (QPE) algorithm, using the indirect measurements of a set of ancillary qubits with a large set of quantum operations to perform in the circuit. 
 This resource demand limited us to only testing the method on quantum emulators, which will probably be usable on quantum platforms after the NISQ period. 
 Alternative methods to restore symmetries, eventually with lower circuit depths and lengths, have been discussed recently in Refs. \cite{Yen19,Lac22}.     
  
One of the promising methods evoked in Ref. \cite{Lac22} is those based on oracles.
Oracles are specific operators that have been introduced in quantum search algorithms.
Among these algorithms, one can mention the Grover method \cite{Gro97a,Gro97b,Kay11,Lim19} that has been recognized as optimal for specific query problems \cite{Zal99,Boy98}.
The practical use of oracles depends strongly on the difficulty of constructing them. We analyze how oracles can be implemented to restore symmetries in the Quantum-Variation After Projection (Q-VAP) method of Ref. \cite{Rui22}.

The procedure reduces the cost of the indirect measurements compared to the QPE to that of a single qubit. It can also continuously monitor symmetry restoration during the variational optimization process.
In particular, it can avoid explicitly projecting the variational state each time the energy is estimated.
Illustrations of the oracle construction are given for the parity, particle number, and total spin symmetries.
Applications are performed on the pairing model.

\section{Quantum Variation-After-Projection with oracle}

Similarly to the Variation-After-Projection (VAP) performed on a classical computer, in the Q-VAP approach, a symmetry-breaking
state $|\Psi\left(\left\{ \theta_{i}\right\} \right)\rangle$ is considered where $\left\{ \theta_{i}\right\}$ are a set of parameters that are 
varied to minimize the energy: 
\begin{equation}
E\left(\left\{ \theta_{i}\right\} \right)=
\frac{\langle\Psi\left(\left\{ \theta_{i}\right\} \right)| \hat H \hat{\mathcal{P}}_{S}|\Psi\left(\left\{ \theta_{i}\right\} \right)\rangle}
{\langle\Psi\left(\left\{ \theta_{i}\right\} \right)|\hat{\mathcal{P}}_{S}|\Psi\left(\left\{ \theta_{i}\right\} \right)\rangle}. 
\label{eq:vap_energy}
\end{equation} 
Here, $\hat H$ denotes the Hamiltonian, and $\hat{\mathcal{P}}_{S}$ is a projector onto the subspace $\hat {\cal H}_S$, 
which respects a given set of symmetries of the Hamiltonian, denoted generically by $S$. In Eq (\ref{eq:vap_energy}), we used the fact that 
$\hat{\mathcal{P}}_{S}^{2}=\hat{\mathcal{P}}_{S}$
and $\left[\hat H, \hat{\mathcal{P}}_{S}\right]=0$. 

The method proposed in Ref \cite{Lac20} to perform symmetry restoration combines the QPE approach and takes advantage of the fact that the eigenvalues of symmetry operators are known. Although this method applies to any symmetry problem, one of its inconveniences is that it relies on two steps to estimate the energy in Eq. (\ref{eq:vap_energy}). 
In the first step, the SR state is obtained by projection, and in the second step, this projected state is used to compute the expectation value of $\hat H$. 
Possible ways to reduce the circuit to perform the projection have been discussed in Ref. \cite{Lac22}. Depending on the projection method, this can lead to quite significant coherence time requirements. 
In most of the techniques discussed in \cite{Lac22}, the projection is achieved by indirect measurements of a set of ancillary qubits. 
Because of this, and despite possible reductions in the number of operations to perform the projection, the first step often remains probabilistic. 
This implies that only part of the events and, depending on the unprojected state properties, a possibly significant fraction of the runs could be thrown away, leading to waste in the use of quantum platforms. 
An extreme situation would be where the SB state has a very small or zero fraction of states belonging to ${\cal H}_S$. 
Thus, the probability of the states with the correct symmetry is too low or zero in the parametric symmetry-breaking wave function. 
In that case, we could find ourselves in a situation where the projector rarely or never projects in the correct subspace. 
To address this issue, we explore below an alternative method where we can directly access the energy in Eq (\ref{eq:vap_energy}) without performing the explicit projection. 

\subsection{Using oracle for symmetry restoration}

An oracle \cite{Gro97a,Gro97b,Lac22}, denoted hereafter by $\hat{O}$, is an operator able to classify the total Hilbert space ${\cal H}$ into two subspaces.
In one subspace, states have a specific property we are interested in. These states are the {\it Good} states and correspond, in our case, to states that respect the symmetry $S$.
The other states are {\it Bad} states and are the ones that do not respect the symmetry. They belong to the complementary subspace of ${\cal H}_S$, denoted hereafter by ${\cal H}_{\bar S}$.

A second property of the oracle is that it acts differently on the good and bad states. In general, the action is relatively simple.
Here, we assume that the oracle multiplies by the same phase $\varphi$ (resp. $\mu$) states that do (resp. do not) respect a symmetry, i.e.:
\begin{equation}
\hat{O}|\psi_k\rangle=\begin{cases}
e^{i\varphi}|\psi_k\rangle & \text{if \ensuremath{|\psi_k\rangle\in{\cal H}_S}}\\
e^{i\mu}|\psi_k\rangle & \text{if \ensuremath{|\psi_k\rangle \in{\cal H}_{\bar S}}}
\end{cases} . \label{eq:definition_oracle}
\end{equation}
We consider below $\varphi \neq \mu$. Note that standard search algorithms usually assume $e^{i\varphi} = -1$ and $e^{i\mu} = + 1$. Here, we define the oracle in a more general way because using phases different from the standard choice will be useful for certain calculations shown below.
The states $\{ | \psi_k \rangle \}$ correspond here to a complete general basis that spans the entire Hilbert space.
Then, the oracle applied to a general wave function $| \Psi \rangle = \sum_k c_k | \psi_k \rangle$ gives:
\begin{eqnarray}
\hat{O}|\Psi\rangle  
 & = & e^{i\varphi}\underbrace{\sum_{\psi_k\in{\cal H}_S} c_{k}|\psi_k\rangle}_{\equiv|\Psi_{G}\rangle}+e^{i\mu}\underbrace{\sum_{\psi_k\notin{\cal H}_S} c_{k}|\psi_k\rangle}_{\equiv|\Psi_{B}\rangle}. \label{eq:goodbad}
\end{eqnarray}
We have that $|\Psi\rangle=|\Psi_{G}\rangle+|\Psi_{B}\rangle$ where $|\Psi_{G}\rangle$ (resp. $|\Psi_{B}\rangle$) corresponds 
to the projection of the state onto the good (resp. bad) subspace.
\begin{figure}[htbp]
\begin{centering}
\begin{tikzpicture}   
    \node[scale=1.25] { 
\begin{quantikz}  
\lstick{ \ket{0}}    & \qw                  & \gate{H} & \ctrl{1}       & \ctrl{1}       & \gate{H} & \meter{}  \\
\lstick{ \ket{\Psi}} & \push{ \qwbundle{n}} & \qw      & \gate{\hat{O}} & \gate{\hat{U}} & \qw      & \qw       
\end{quantikz} 
};  
\end{tikzpicture}\\ 
\par\end{centering}
\begin{centering}
\caption{Hadamard test  to get the expectation value of the operator $\hat{U}$ in a determined subspace defined by $\hat{O}$ (see text and Eq. (\ref{eq:definition_oracle})).
$\hat{O}$ stands for the oracle operator. In this circuit, one ancillary qubit is used.  
The repeated measurements of $0$ and $1$ of the ancillary qubit give access to the real part (or to the imaginary part if a phase gate of phase $-\pi/2$ is 
added after the first Hadamard gate on the ancilla qubit) of the expectation value given by Eq. (\ref{eq:expfinal}).  
\label{fig:hadamard_oracle} }
\par\end{centering}
\end{figure}
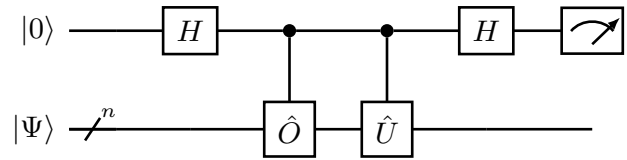

We then perform a Hadamard test using a generic operator $\hat U$ and the oracle. 
The corresponding circuit is shown in Fig \ref{fig:hadamard_oracle}. 
The choice of the operator $\hat U$ is discussed below. 
The Hadamard test gives access to the following expectation value:
\begin{eqnarray}
\langle\Psi|\hat{U}\hat{O}|\Psi\rangle&=&\left(\langle\Psi_{G}|+\langle\Psi_{B}|\right)\hat{U}\left(e^{i\varphi}|\Psi_{G}\rangle+e^{i\mu}|\Psi_{B}\rangle\right) ,
\nonumber \\
&=& e^{i\varphi}\left(\langle\Psi_{G}|\hat{U}|\Psi_{G}\rangle+\langle\Psi_{B}|\hat{U}|\Psi_{G}\rangle\right) \nonumber \\
&+&e^{i\mu}\left(\langle\Psi_{G}|\hat{U}|\Psi_{B}\rangle+\langle\Psi_{B}|\hat{U}|\Psi_{B}\rangle\right) . \label{eq:expfinal}
\end{eqnarray}
This expression simplifies if $\hat{U}$ also preserves the symmetry $S$, so that we have 
$\hat{U}|\Psi_{G}\rangle\in{\cal H}_S$ and $\hat{U}|\Psi_{B}\rangle\in{\cal H}_{\bar S}$. Accordingly, we deduce 
\begin{eqnarray}
\langle\Psi_{B}|\hat{U}|\Psi_{G}\rangle & =&\langle\Psi_{G}|\hat{U}|\Psi_{B}\rangle =0.
\end{eqnarray}
With the use of the Hadamard test, we can retrieve the real or imaginary part of the expectation value:
\begin{eqnarray}
\langle\Psi|\hat{U}\hat{O}|\Psi\rangle&=& e^{i\varphi}\langle\Psi_{G}|\hat{U}|\Psi_{G}\rangle+e^{i\mu}\langle\Psi_{B}|\hat{U}|\Psi_{B}\rangle. \label{eq:result_hadamard_oracle}
\end{eqnarray}

An interesting situation is the case where $\hat{U}$ is Hermitian and where we set  $\varphi=0$ and
$\mu=\pi/2$. Then, the Hadamard test used for obtaining the real part gives the expectation value of the operator $\hat{U}$ on the 
state of interest, i.e. 
\begin{eqnarray}
p_0 - p_1 = \langle\Psi_{G}|\hat{U}|\Psi_{G}\rangle. \label{eq:hadamard_optimal}
\end{eqnarray}
where $p_0$ (resp. $p_1$) denotes the probability to measure $0$ or $1$ in the ancillary qubit. 
Therefore, we see that the method provides direct access to the expectation value of any operator $\hat U$
taken on the ``Good'' state, i.e., symmetry restored part of the initial wave function. 
 
\begin{figure}[htbp]
\includegraphics[scale=0.5]{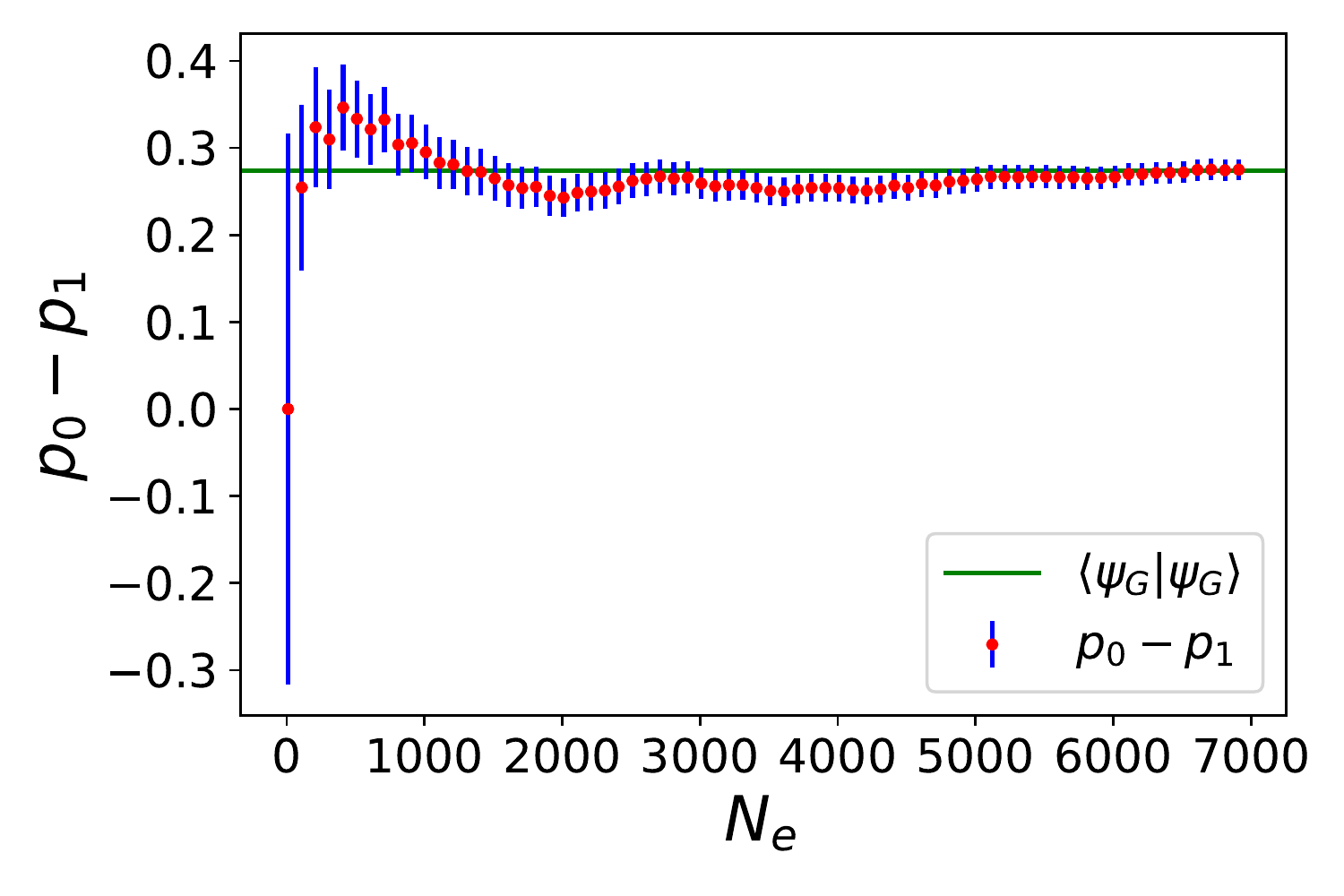}  
\caption{
Illustration of the use of an oracle to obtain $\langle\Psi_{G}|\hat{U}|\Psi_{G}\rangle$ for the specific case $\hat U =I$ and $(\varphi=0, \mu=\pi/2)$ in which case the oracle gives the amplitude of the initial state belonging to ${\cal H}_G$. The initial state is encoded here on $8$ qubits, and it corresponds to
the equiprobable state $|\Psi\rangle = \frac{1}{2^4}\sum_k |k\rangle $ where $| k \rangle$ should be interpreted as the binary representation of the integer $k$ encoded on the $8$ register qubits. In this example, the "Good" states were defined as the states with four $1$ in their binary representation. Suppose the initial state represents a many-body system encoded on the quantum register using the Jordan-Wigner fermion to qubit mapping  \cite{Jor28,Lie61,Som02,Fan19}. Then, the projected
state will correspond to a many-body system with exactly $4$ particles while the initial state mixes all possible particle numbers. The green horizontal line is the exact value of $\langle\Psi_{G}|\Psi_{G}\rangle$, while the red dots correspond to the value $p_0 - p_1$ deduced from a set of $N_e$ measurements. The error bars in blue shown in the figure are taken as $1/\sqrt{N_e}$. The oracle is constructed using the method discussed in section \ref{sec:oracleconstruction}.
}
\label{fig:p_events}
\end{figure}

An illustration of the use of Eq. (\ref{eq:hadamard_optimal}) using the
probability $p_0$ and $p_1$ with a limited set of measurements $N_{e}$ is shown in Fig. \ref{fig:p_events}.
This figure illustrates that an approximate value of the left-hand side of Eq. (\ref{eq:hadamard_optimal}) can be 
obtained even if a limited number of measurements is performed. An important aspect is that no event is rejected
contrary to methods usually based on the iterative use of Hadamard tests (see for instance discussion in \cite{Lac22}).

\begin{figure}[htbp]
\includegraphics[scale=0.6]{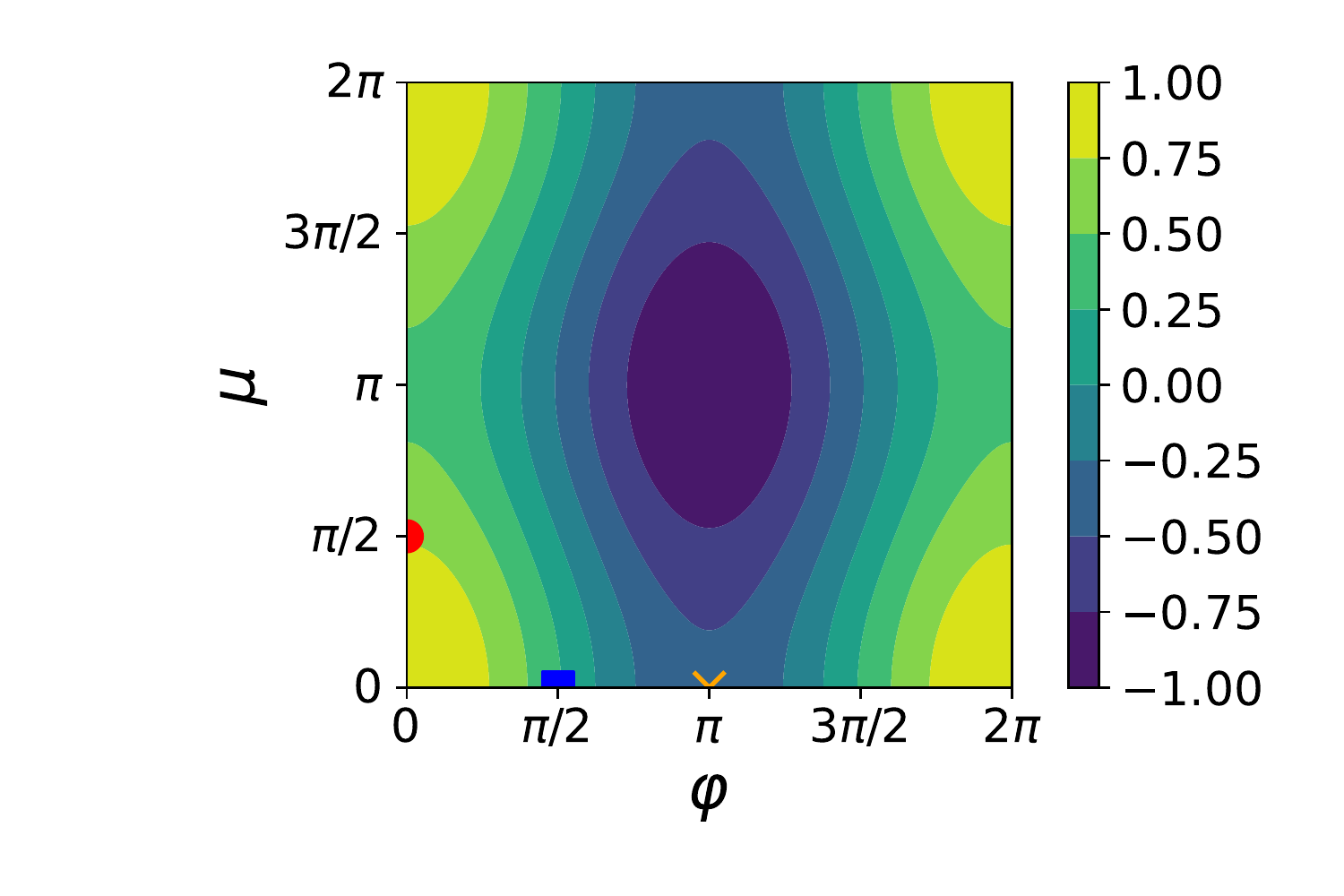}
\caption{Contour plot of the ${\rm Re}[ \langle \Psi | \hat{O} (\varphi, \mu ) |\Psi \rangle ]$ quantities deduced from Eq. (\ref{eq:expfinal})
as function of the angles $\varphi$ and  $ \mu$ $\in \left[0, 2\pi \right]$ 
for the same initial state and oracle considered in Fig. \ref{fig:p_events}. The  symbols highlight specific angles: 
the angles used in Fig.  \ref{fig:p_events} [red filled circles], the case $(\varphi= \pi/2, \mu=0)$ that leads to 
$p_0 - p_1 = \langle \Phi_B| \Phi_B \rangle$ [blue filled square] and the standard prescription for the oracle generally used in Grover search algorithm, i.e $(\varphi = \pi, \mu=0)$ [orange cross]. }
\label{fig:energie_oracle}
\end{figure}

We also note in passing that the Eq. (\ref{eq:result_hadamard_oracle}) is a rather simple trigonometric function of the angles $(\varphi, \mu)$. 
It only depends on the two parameters $\langle\Psi_{G}|\hat{U}|\Psi_{G}\rangle$ and $\langle \Psi_{B}|\hat{U}|\Psi_{B}\rangle$. 
Therefore, we see that the knowledge of $\langle\Psi|\hat{U}\hat{O}|\Psi\rangle$ given by Eq. (\ref{eq:expfinal}) for few selected values of $(\varphi, \pi)$ 
give access to the same expectation values for all values of these angles. For the case presented in Fig. \ref{fig:p_events} for $\hat U=I$, a single pair of angles is sufficient since:
\begin{eqnarray}
\langle\Psi_{G}|\Psi_{G}\rangle+\langle\Psi_{B}|\Psi_{B}\rangle = 1. \nonumber 
\end{eqnarray}
The corresponding function ${\rm Re}[ \langle \Psi | \hat O(\varphi, \mu) | \Psi \rangle ]$ is illustrated in Fig.  \ref{fig:energie_oracle} as well as
few remarkable sets of angles. 

In addition to the advantage, compared to the method proposed in \cite{Lac20,Rui22}, that the circuit depth is reduced by using a single ancillary qubit, 
we see that repeated measurements of this qubit will rapidly enable us to identify if the initial state has a significant amplitude belonging to the space ${\cal H}_S$. 
In the opposite case, i.e., if $p_0 \simeq p_1$, the initial state can be monitored to increase this amplitude variationally. 
Last, we mention that there are no wasted events in the approach, contrary to most of the methods discussed in \cite{Lac22}.      

Up to now, we have not discussed the operator $\hat U$ itself. 
We will typically be interested in using $\hat U = I$ or  $\hat U = \hat H$ for many-body systems. 
The former gives access to the norm of the state $\langle\Psi_{G}|\Psi_{G}\rangle$ while the latter provides the expectation value of the Hamiltonian
over the ``Good'' space. The ratio of the two quantities gives the projected energy (\ref{eq:vap_energy}) for the symmetry-restored state. 

Since  $\hat H$ is hermitian but not unitary, it might be helpful to use $\hat{U}(t)=e^{-it \hat H}$ instead of $\hat H$ itself. 
We can extract the energy using the generating function
method discussed in details in Ref. \cite{Rui21}, i.e.:
\begin{eqnarray}
\langle\Psi_{G}|\Psi_{G}\rangle= F_{G}(0) , ~~
\langle\Psi_{G}|\hat H|\Psi_{G}\rangle=i\left.\frac{dF_{G}(t)}{dt}\right|_{t=0}, \label{eq:generating_function} 
\end{eqnarray}
where $F_{G}(t)=\langle\Psi_{G}|e^{-it \hat H}|\Psi_{G}\rangle$ is obtained using the oracle technique. 
Given that, in this case, we would be interested in the imaginary part of $F_{G}(t)$, 
we can use a modified Hadamard test to get the imaginary part of the expected value (see Fig. 1 of ref. \cite{Rui21}). 
In practice, the first derivative of $F_G(t)$ can be evaluated using a simple finite-difference method, 
in which case only short-time propagation (short circuit length) is required. 
Alternatively, to reduce the numerical noise of this derivative, one can use the parameter-shift rule technique \cite{Mit18, Sch19}. 
 
\section{Oracle construction for symmetry-restoration}
\label{sec:oracleconstruction}

The approach presented above strongly relies on the possibility of efficiently constructing the oracle associated with the symmetry restoration problem. 
A SB problem's specificity is that the eigenvalues of the symmetry operator $\hat S$ are usually known. 
We restrict the discussion here to the case where the eigenvalues are discrete. 

Because the number of qubits $n_q$ that describe the system is finite in practice, 
these eigenvalues are bound from below and above. 
We denote them by $\lambda_1< \cdots < \lambda_{\Omega}$ where $\Omega$ depends on $n_q$.  In the following, we define the set of values 
$\xi_\alpha = \lambda_\alpha - \lambda_1$ and assume that we can write $\xi_\alpha = a m_\alpha$ 
where $a$ is a constant that depends on the symmetry considered and $\{ m_\alpha \}_{\alpha=1,\Omega}$ 
correspond to a set of positive integers in the interval $\left[0, m_\Omega\right]$. 
The projector $\hat P_\alpha$ that projects onto the subspace associated with the eigenvalue $\lambda_\alpha$ can then be written as:

\begin{equation}
\hat{P}_{\alpha} = \sum_{k=0}^{M} \alpha_{k} e^{i\phi_{k}\hat{S}} , \label{eq:reduced_projector_nb_particles} 
\end{equation}
with 
\footnote{The proof can be made using simply the identity:
\begin{eqnarray}
\frac{1}{M+1} \sum_{k=0}^M e^{2i\pi \frac{k (m-n)}{M+1}} = \delta_{nm}  
\end{eqnarray}
that is correct provided that $m,n \in \left[0,M\right]$. }
\begin{eqnarray}
\left\{ 
\begin{array}{l}
\displaystyle \alpha_k = \frac{1}{M+1} e^{-i\phi_k \left(\xi_\alpha + \lambda_1\right)}   \\
\\
\phi_k  = \frac{2\pi k}{a (M+1)}
\end{array}
\right. \label{eq:alphaphi}
,
\end{eqnarray}
and $M=m_{\Omega}$. This generic form of a projector is particularly interesting for quantum computing since the non-unitary projector 
is written as a linear combination of unitary (LCU) operators. 
This form could be advantageously used to reduce the cost of performing expectation values of operators. 
For instance, an operator's projected expectation value of an operator $\hat A$ verifies:
\begin{eqnarray}
\langle \hat A \hat P_\alpha \rangle = \sum_k \alpha_k \langle \hat A e^{i \phi_k \hat{S}} \rangle. \label{eq:projection_decompositon}
\end{eqnarray}     
Each expectation value on the right-hand side can be separately obtained in a quantum computer, and the sum can be reconstructed on a classical computer. 
An illustration of this technique can be found in Ref. \cite{Kha21}. 
The possible advantages of using the LCU decomposition will be discussed below in more detail.  

As underlined in Ref. \cite{Lac22}, once we have a usable form for a projector, 
one can deduce a similar form for the oracle. 
The oracle operator, as defined in Eq (\ref{eq:definition_oracle}), can be constructed from the projector as:
\begin{eqnarray}
\hat{O}_{\alpha}&=& e^{i\mu}\left(\hat{I}-\hat{P}_{\alpha }\right)+e^{i\varphi}\hat{P}_{\alpha}. \label{eq:definition_oracle_nb_particles}
\end{eqnarray}
Using Eq (\ref{eq:reduced_projector_nb_particles}), we get:
\begin{eqnarray}
\hat{O}_{\alpha}  = e^{i\mu}\hat{I}+\left(e^{i\varphi}-e^{i\mu}\right)\hat{P}_{\alpha} \equiv \sum_{k=0}^{M} \beta_k(\phi, \varphi) e^{i\phi_{k}\hat{S}} \label{eq:oracle}
\label{eq:lcuoracle}
\end{eqnarray}
with 
\begin{eqnarray}
\left\{ 
\begin{array}{l}
\beta_0 = e^{i\mu}+ (e^{i\varphi}-e^{i\mu}) \alpha_0  \\
\\
\beta_{k \neq 0} = (e^{i\varphi}-e^{i\mu}) \alpha_k
\end{array}
\right. \label{eq:beta}
\end{eqnarray}

If an operator $\hat A$ respects the symmetry, we can 
show that its expectation value on a projected normalized state verifies:  
\begin{eqnarray}
\frac{\langle  \hat{A} \hat{P}_{\alpha} \rangle}{\langle\hat{P}_{\alpha} \rangle} &=&
\frac{\langle \hat{A} \hat O_\alpha \rangle - e^{i \mu} \langle \hat{A}  \rangle}{\langle\hat{O}_{\alpha} \rangle - e^{i \mu}}, 
  \label{eq:projA}
\end{eqnarray}
where we used the compact notation $\langle \Psi | .| \Psi \rangle = \langle . \rangle$ and assumed that the state $| \Psi \rangle$ is normalized to $1$. 
It is interesting to note that the above equality holds whatever is the retained value of $\varphi$ and $\mu$ provided that $\varphi \neq \mu$. 
This flexibility could be used to reduce the numerical cost. 
We mention the special case where $\hat A$ is a hermitian operator. In this case, we can set $\varphi=0$ and $\mu=\pi/2$, 
and we would only need to compute the real part of $\langle \hat{A} \hat{O} \rangle$ and $\langle \hat{O} \rangle$.
 
 
\subsection{Example of projectors and oracle for selected symmetries}

We consider here some example of common symmetries that are standardly 
respected by many-body hamiltonians.   \\

\noindent {\bf Parity}: 
One of the simplest symmetries is the one associated with parity; its operator is denoted hereafter as $\hat \pi$.
We consider below a set of computational states $| m \rangle$ with $m=0, 2^{n_q} - 1$ mapped to a qubit register
 with $n_q$ qubits such that $|m\rangle = | \delta_{n_q - 1}, \cdots , \delta_0 \rangle$ with $m = \sum_{i=0}^{n_q-1} \delta_j 2^{j}$.
Such states are eigenstates of $\hat \pi$ with eigenvalues $+1$ (resp. $-1$) if the number of $\delta$'s equal to $1$ is even (resp. odd).
For many-body Fermi systems, if one uses the Jordan-Wigner Transformation (JWT) to map the Fock space to the qubit space,
 \cite{Jor28,Lie61,Som02,Fan19}, the occupation probability of
a particle in a particular state becomes equivalent to the probability of measuring the state $|1\rangle$ in the corresponding qubit.
Then the qubit parity identifies with the many-body state's parity and separates the total Hilbert space into odd and even particle number subspaces.
For the special case $(\varphi, \mu)=(0, \pi)$ in Eq. (\ref{eq:definition_oracle}), the parity case is very peculiar because the parity operator is itself the oracle $\hat O_\pi = \hat \pi$.
In the following, we use the notation $(X_m, Y_m, Z_m)$ for the Pauli matrices acting on the $m^{th}$ qubit.
The parity operator can be written as $\hat \pi = \bigotimes_{m=0}^{n_q-1} Z_m$;
 this operator was used in \cite{McA19} to mitigate the error in quantum computers, and a low-cost method, similar to Eq. (\ref{eq:projA}) was proposed in Ref. \cite{Bon18,Sag19} to get the parity projected densities.
Denoting by $\varepsilon_\pi = \pm 1$ the eigenvalues of the parity operator, the projector is given by:
\begin{eqnarray}
P_{\varepsilon_\pi} = \frac{1}{2} \left( I + \varepsilon_\pi \hat \pi  \right).
\end{eqnarray}      
The generalized form of the oracle for the parity operator as given in Eq. (\ref{eq:definition_oracle}) is:
\begin{eqnarray}
O_{\varepsilon_\pi} = \frac{1}{2} \left( e^{i\varphi} + e^{i\mu} \right) I + \frac{1}{2} \left( e^{i\varphi} - e^{i\mu} \right) \varepsilon_\pi \hat \pi  . 
\end{eqnarray}

\noindent {\bf Particle number:} 
If we use again the JWT encoding for fermions, the particle number operator $\hat N$ is given by $\hat N =\sum_{i=0}^{n_q - 1} \left( I_i - Z_i \right)/2$.
States of the computational basis are again eigenstates of this operator with eigenvalues $N=0, \cdots, n_q$ that correspond 
to the number of $1$ in the binary representation of the states. Projection on particle number with the QPE approach was illustrated in ref. \cite{Lac20,Rui22}.  
For this operator, the decomposition (\ref{eq:reduced_projector_nb_particles}) for the projection $\hat P_A$ onto a given particle number $A$  
is obtained by using $a=1$, $\lambda_1 = 0$ and $m_\alpha = A$, while $M$ should verify $M \geq n_q$ in Eq. (\ref{eq:alphaphi}). The different unitary operators 
entering in Eq. (\ref{eq:reduced_projector_nb_particles}) can be implemented using the tensor decomposition \cite{Rui21}:
\begin{equation}
e^{i\phi_{k}\hat{N}}=\bigotimes_{m=0}^{n_q -1} 
\begin{pmatrix}1 & 0\\
0 & e^{i\phi_{k}}
\end{pmatrix}_m 
,\label{eq:particle_number_operator_to_q_gates}
\end{equation}
where the set of $2 \times 2$ matrix on the right side corresponds to phase gates acting on the set of qubits.  

\noindent {\bf Total spin operator:} We now consider a slightly more complex situation where the qubit register corresponds to a set of spins. 
We use the same mapping from spins to qubits as in Ref \cite{Siw21}, and we consider $n_q$ spins encoded on a $n_q$ qubits register 
with the convention $\left\{ |0\rangle_{m},|1\rangle_{m}\right\} _{m=0,\dots,n_q-1}=\left\{ |\uparrow\rangle_{m},|\downarrow\rangle_{m}\right\} _{m=0, \dots, n_q-1}$. 
The spin vector for the qubit \emph{m} is directly denoted as $\vec{S}_{m}=\frac{1}{2}\left(X_{m},Y_{m},Z_{m}\right)$. 
The QPE method was employed in \cite{Siw21} to project a state simultaneously onto a given value of the total spin and its $z$-component. 
Hereafter, we denote by $S(S+1)$ and $S_z$ the corresponding eigenvalues (note that here we use the convention $\hbar =1$). 
As shown in Ref. \cite{Siw21}, the projection on a given eigenstate of $\hat S_z$ is equivalent to the projection on particle $\hat N$ discussed above. 
Therefore, we only discuss the projection on eigenstates of $\hat {\bf S}^2$. It is first helpful to write this operator as \cite{Low69}:
\begin{equation}
\hat {\bf S}^{2}=\frac{n_q\left(4-n_q\right)}{4}+\sum_{i<j,j=0}^{n_q-1} \hat P_{ij}, \label{eq:s_2}
\end{equation}
where $\hat P_{ij}=\frac{1}{2}\left(I_{i}I_{j}+X_{i}X_{j}+Y_{i}Y_{j}+Z_{i}Z_{j}\right)$ are the permutation operators that invert two qubits values, i.e. $\hat P_{ij} | \delta_i \delta_j \rangle = | \delta_j \delta_i \rangle$. 

The case of $\hat {\bf S}^2$ slightly differs from the particle number since this operator has non-equidistant eigenvalues $S(S+1)$. 
Its possible values depend on the number of qubits $n_q$; more precisely, we have:

\begin{itemize}
  \item For even $n_q$, we have $S=0,1, \cdots, n_q / 2$ leading to $S(S+1) = 0,2, \cdots, n_q (n_q+2) /4$; from which we deduce $\lambda_1 = 0$. 
  Then to reduce the number of phase operators in Eq. (\ref{eq:reduced_projector_nb_particles}) one can assume $a=2$, so that $m_\alpha=0,1, \cdots, k_q (k_q +1)/2$ with $k_q = n_q/2$.   
  \item For odd  $n_q$, we have $S=\frac{1}{2}, \frac{3}{2}, \cdots, \frac{n_q}{2}$ leading to $\lambda_1 = \frac{3}{4}$ and 
  $S(S+1) - \lambda_1 = 0, 3 , 8 , \cdots, k_q (k_q + 2)$ with $n_q = 2 k_q + 1$. Given this sequence, $a$
  should be set to $1$ and the possible values of $m_\alpha$ directly identifies with those reported for $S(S+1) - \lambda_1$. 
\end{itemize} 

To implement the circuit associated with the different phase operators $e^{i \phi_k \hat {\bf S^2} } $ we directly used here Eq. (\ref{eq:s_2})
together with the Trotter-Suzuki method \cite{Tro59,McA20}. Technical aspects and related quantum circuits are given in appendix \ref{app:s2}. 
 We note that for the total spin, we can also use the method of Ref. \cite{Tsu20}, where 
the discretization of the projector is written as an integral over spherical coordinate angles. The latter method gives an alternative approximate 
LCU decomposition of the total spin projector with the advantage of only needing one-body operators in each exponential term.

\subsection{Practical aspects for the oracle application}  

In the above discussion, we have shown first how the oracle can be used for restoring symmetries and how it can be decomposed systematically in the LCU form using Eq. (\ref{eq:reduced_projector_nb_particles}).
Here, we discuss in more detail how, starting from the LCU decomposition of oracles, they can be implemented on quantum computers.
Two methods are presented, one that can be readily applied on NISQ platforms and the second based on the LCU algorithm \cite{Lon06}.

\subsubsection{Implementation of oracle on NISQ devices}

One standard approach to reduce the numerical cost associated with an operator 
is to separate its expectation values into a set of independent evaluations requiring a lower cost.  
From Eq. (\ref{eq:projA}), we see that the expectation value of an operator $\hat A$ 
on the projected state can already be decomposed either directly as $\langle AP_{\alpha}\rangle =\sum_k \alpha_k \langle Ae^{i\phi\hat{S}}\rangle $
and $\langle P_{\alpha} \rangle = \sum_k \alpha_k \langle e^{i\phi_k\hat{S}}\rangle $ or into the expectation values of 
$\langle \hat{A} \rangle$, $\langle\hat{O}_{\alpha} \rangle$, $\langle \hat{A} \hat O_\alpha \rangle $; 
this illustrates that the projected state itself does not need to be explicitly constructed. 

Provided that the operator $\hat A$ can also be written as an LCU such that $\hat A = \sum_j \gamma_j \hat A_j$, where all 
$\hat A_j$ are unitary operators, we deduce:
\begin{eqnarray}
\langle \hat{A} \rangle &=& \sum_j \gamma_j \langle \hat{A}_j  \rangle,~~
\langle \hat{O}_{\alpha} \rangle = \sum_k \beta_k \langle \hat{V}_k \rangle, \nonumber \\
\langle \hat{O}_{\alpha} \hat{A}\rangle &=& \sum_{k,j} \beta_k \gamma_j \langle \hat{V}_k \hat A_j \rangle, \nonumber 
\end{eqnarray}   
where we used Eq. (\ref{eq:lcuoracle}) with the notations $\hat V_k = e^{i\phi_{k}\hat{S}}$ and $\beta_k = \beta_k(\phi, \varphi) $.
A standard pragmatic approach to reduce the cost of the evaluation of these expected values is to use a Hadamard test
to evaluate each expectation among the set $\{ \langle \hat{A}_j \rangle , \langle \hat{V}_k \rangle , \langle \hat{V}_k \hat A_j \rangle \}$.
We also mention that given the appropriate decomposition of $\hat{A}$ and $\hat{O}$,
 it is possible to use classical shadows \cite{Hua20} to reduce even more the amount of QC resources used.
It is worth noting that such expectation values are independent of the choice $(\mu, \varphi)$ used in the general definition
 (\ref{eq:definition_oracle}) and can be made only once to get access to the whole family of operators $\hat O_{\alpha} (\mu, \varphi)$.
  An illustration of this aspect is given in Fig. \ref{fig:energie_oracle}.


\subsubsection{Oracle or projector construction with a simplified LCU algorithm}

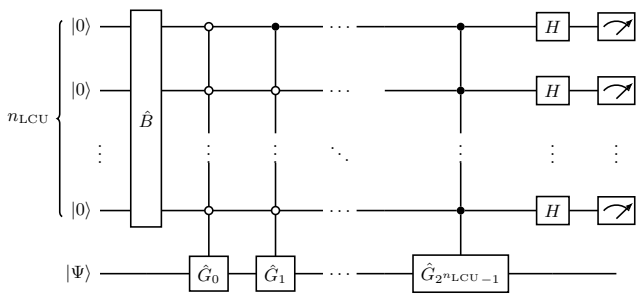
\begin{figure}[htbp]
\begin{centering}
\begin{tikzpicture}   
    \node[scale=0.75] { 
\begin{quantikz}  
\lstick[wires=4]{$n_{\rm LCU}$} &\lstick{  \ket{0}}   & \gate[wires=4, nwires={3}]{\hat{B}} & \octrl{1}           & \ctrl{1}           & \ \ldots\ \qw & \qw               & \ctrl{1}                  & \gate{H} & \meter{} & \\
                            &\lstick{ \ket{0}}    &                                     & \octrl{1}           & \octrl{1}          & \ \ldots\ \qw &                   & \ctrl{1}                  & \gate{H} & \meter{} & \\
                            &\vdots               &                                     & \vdots              & \vdots             & \ddots        &                   & \vdots                    & \vdots   & \vdots   & \\   
                            &\lstick{ \ket{0}}    &                                     & \octrl{1} \vqw{-1}  & \octrl{1} \vqw{-1} & \ \ldots\ \qw & \qw               & \ctrl{1} \vqw{-1}         & \gate{H} & \meter{} & \\   
                            &\lstick{ \ket{\Psi}} & \qw                                 & \gate{\hat G_{0}}        & \gate{\hat G_{1}}       & \ \ldots\ \qw & \qw               & \gate{\hat G_{2^{n_{\rm LCU}}- 1}} & \qw      & \qw      & 
\end{quantikz} 
}; 
\end{tikzpicture}\\
\par\end{centering}
\centering{}\caption{Simplified LCU circuit with $n_{\rm LCU}$ ancilla qubits. This circuit
can implement a linear combination of up to $2^{n_{\rm LCU}}$ unitary
operators. The filled circles (resp. open) are controlled operations by the $|1\rangle$ (resp. $|0\rangle$)
state. We use the Qiskit \cite{Ibm21} convention in all the text, where the uppermost qubit corresponds to the least significant bit in the binary representation.
\label{fig:simplified_LCU_circuit}}
\end{figure}

We now discuss a general approach to applying a non-unitary operator written as an LCU on a quantum computer.
Given an operator written as a linear combination of unitary operators $\hat G = \sum_{k=0}^{k_{\rm max} -1} g_k \hat G_k$, the LCU
algorithm \cite{Lon06} gives a method to perform the following operation:
\begin{eqnarray}
| \Psi \rangle \longrightarrow \hat G | \Psi \rangle. \label{eq:multnonh}
\end{eqnarray}
This method has been extensively used for Hamiltonian simulation \cite{Chi12,Ber14,Ber15}. 
The LCU method might be somewhat demanding regarding ancillary qubits 
to be added to the system description and the number of operations to implement it. 
Nevertheless, it remains an excellent candidate for implementing projectors 
for symmetry-restoration and associated oracles in quantum computers beyond the NISQ period. 
We use the implementation proposed in \cite{Wei20}, shown in Fig.  \ref{fig:simplified_LCU_circuit}, to do the operation (\ref{eq:multnonh}). 
We need to introduce a set of $n_{\rm LCU}$ ancillary qubits such that $2^{\rm LCU} \ge k_{\rm max}$. 
Then, an operator $\hat{B}$  acting on the ancillary qubits is defined as:
\begin{align}
\hat{B}|0\rangle^{\otimes n_{LCU}} & =\frac{1}{\mathcal{N}}\sum_{k=0}^{2^{n_{LCU}} - 1} g_{k}|k\rangle_{\rm LCU} \label{eq:b} 
\end{align}
with $g_{k \geq k_{max}} = 0$ and where we introduced the normalization constant $\mathcal{N}=\sqrt{\sum_{k} |g_{k}|^2}$.
Here $\hat G$ can be any of the operators discussed above, e.g. 
the projector (\ref{eq:reduced_projector_nb_particles}) or the associated oracle (\ref{eq:oracle}). 

After applying the operation (\ref{eq:b}) to the ancillary register, a set of states $\hat G_{k} |\Psi\rangle$ 
are sequentially associated with the different computational states of the ancillary register using 
controlled operations as shown in Fig. \ref{fig:simplified_LCU_circuit}. 
Finally, the operation $H^{\otimes n_{LCU}}$ is applied to the
ancillary register. By measuring the qubits in the ancillary register and selecting the events where the state 
$|0\rangle ^{\otimes n_{LCU}}$ is measured, we obtain that the system state, after each of these measurements, 
collapses to the desired $\hat G | \Psi \rangle$ state.  

For the parity or particle number projection, the operators $\hat G_k$ as defined in Eq. (\ref{eq:particle_number_operator_to_q_gates})  are diagonal in the computational basis.
A general implementation of a diagonal operator that can be applied in the case of these two types of oracles is presented in Refs. \cite{Bul03,Wel14}.
These works showed that the optimal gate count for an arbitrary diagonal operator was $2^{n_q+1}-3$ where $n_q$ is the number of qubits.
However, for these cases, each operator $\hat{G}_k$ controlled by the $n_{LCU}$ qubits in the ancilla register 
can be implemented using a linear number of gates in $n_{LCU}$ and $n_q$. This linear dependence in
$n_{LCU}$ originates from the fact that a single qubit unitary gate controlled by $n$ qubits can be decomposed in single qubits gates and CNOTS whose number depends linearly in $n$ \cite{Sil22}.
 
The use of the set of Hadamard gates $H^{\otimes n_{LCU}}$ in Fig \ref{fig:simplified_LCU_circuit} leads to an exponential decrease in
the probability of measuring precisely zero in all ancillary qubits, i.e. $p_{|0\rangle^{\otimes n_{LCU}} } \sim 1/\sqrt{2^{n_{LCU}}}$, which 
leads to a large set of rejected events.  
To avoid this unwanted feature, we replace the set of Hadamard gates with an operator $\hat{E}^{\dagger}$, where $\hat{E}$ is the 
operator that initialize an equiprobable distribution in the ancillary register up to the state $k_{max}$, i.e.:
\begin{equation}
\hat{E} |0\rangle^{\otimes n_{LCU}}=\frac{1}{\sqrt{k_{max}}}\sum_{k=0}^{k_{max}-1}|k\rangle_{\rm LCU} .\label{eq:equiprobable_operator}
\end{equation}

This replacement can be done because the only requirement for an arbitrary operator $\hat{M}$ to be able to replace the set of Hadamard gates is to have 
equal components in its $M_{0, j \leq k_{max}}$ entries. Using $\hat{M} = \hat{E}^{\dagger}$ defined in Eq. (\ref{eq:equiprobable_operator}), 
the probability $p_{|0\rangle^{\otimes n_{LCU}} }$ now scales as $\sim 1/\sqrt{k_{max}}$ and leads to a significant reduction of the number of rejected events compared to the 
initial proposal shown in Fig. \ref{fig:simplified_LCU_circuit}.
We mention that the construction of such an operator is polynomial in the number of qubits $\mathcal{O}\left(Ln_{LCU}^{2}\right)$\cite{Lon20}. 

\section{Example of applications}

\begin{figure}[htbp]
\begin{centering}
\includegraphics[scale=0.45, trim = 0.5cm 0.2cm 0.5cm 0.5cm]{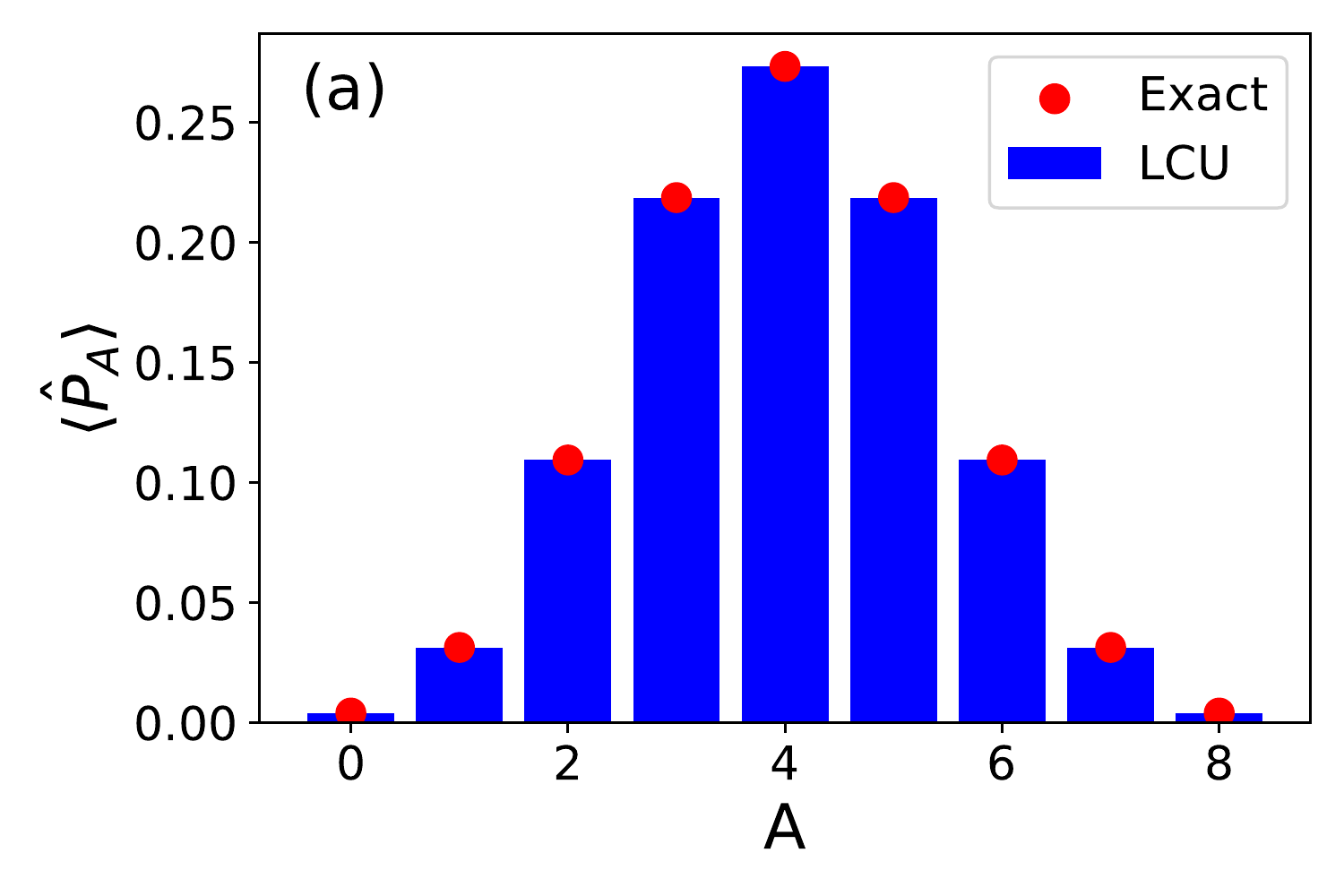}
\includegraphics[scale=0.45, trim = 0.5cm 0.6cm 0.5cm 0.5cm]{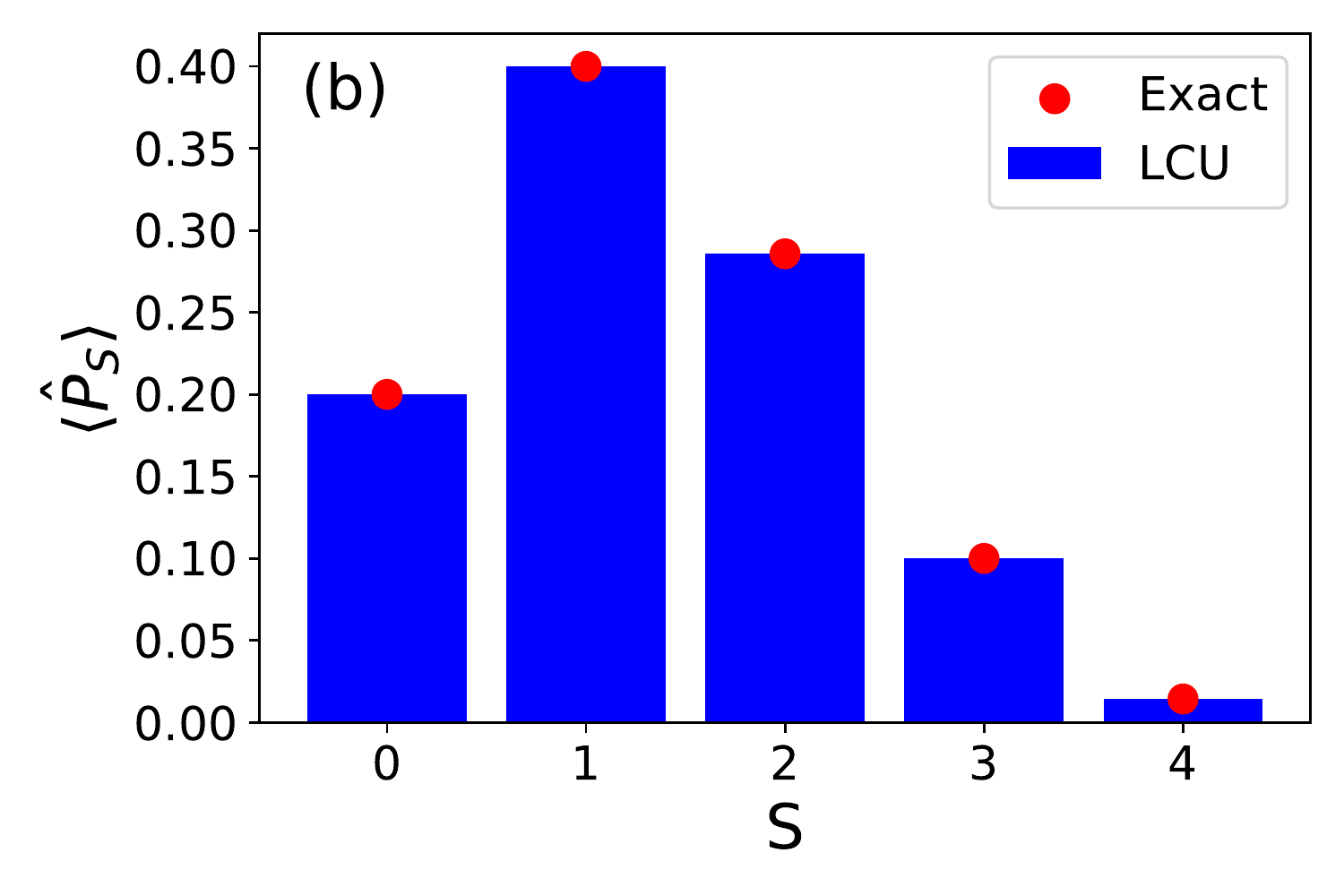}
\par \end{centering}
\caption{Example of projections performed with the oracle technique for (a) the number of particles  and (b) total spin. In panel (a), the same initial state as 
in Fig. \ref{fig:p_events} and \ref{fig:energie_oracle} is used. 
For the spin case, we consider an initial state encoded on $8$ qubits and 
written as $|\psi \rangle = \bigotimes_{k=0}^{n_q - 1} H_k \bigotimes_{k'=0}^{3} X_{k'} |0\rangle ^{\otimes n_q}$.  
In both panels, the red-filled circles correspond to the exact results, while the blue histograms are the results obtained 
by performing the projection through the oracle method. }
\label{fig:part-spin}
\end{figure}

Some illustrations of applications of the oracle techniques for projection as an alternative to the previously proposed methods of \cite{Lac20,Siw21,Rui22} are given in this section.
In Fig \ref{fig:part-spin}, we show examples of how by only computing the expected values $\langle \hat{A}e^{i\phi_k \hat{S}} \rangle $ from Eq. (\ref{eq:projection_decompositon})
we have access to all the projected expectation values over the ensemble of eigenvalues of the symmetry operator. For the particle number symmetry, we use
the same initial state as the one used in Figs. \ref{fig:p_events} and \ref{fig:energie_oracle} that corresponds to an equiprobable mixing of all possible states of the total qubit space basis. 
Assuming the Jordan-Wigner mapping, these states mix different particle numbers.
We show in panel (a) of Fig. \ref{fig:part-spin}, the amplitudes of the initial state obtained for different particle number $A=0,\cdots,8$.
The simulations have been obtained using the LCU decomposition of the operator $\hat P_A$ with varying particle numbers. All the simulations were made using the Qiskit package \cite{Ibm21}.

A second illustration is given in panel (b) of Fig. \ref{fig:part-spin} where the projection on the total spin is performed for different values 
of $S$. In this case, similarly to the encoding of spins used in Ref. \cite{Siw21} and discussed previously, 
the spin components $(| \uparrow \rangle , | \downarrow \rangle)$ of a particle is directly mapped to the states $(|0 \rangle, | 1 \rangle )$ 
of a given qubit. Accordingly, possible values of $S$ ranges from $0$ to $4$.  Since here we performed 
the calculation onto a quantum computer emulator, i.e.  without noise, we deduce that the amplitude decompositions matches 
 the exact amplitudes. 
 
\begin{figure}[htbp]
\begin{center}
\includegraphics[scale=0.5, trim = 0.5cm 0.3cm 0.4cm 0.5cm]{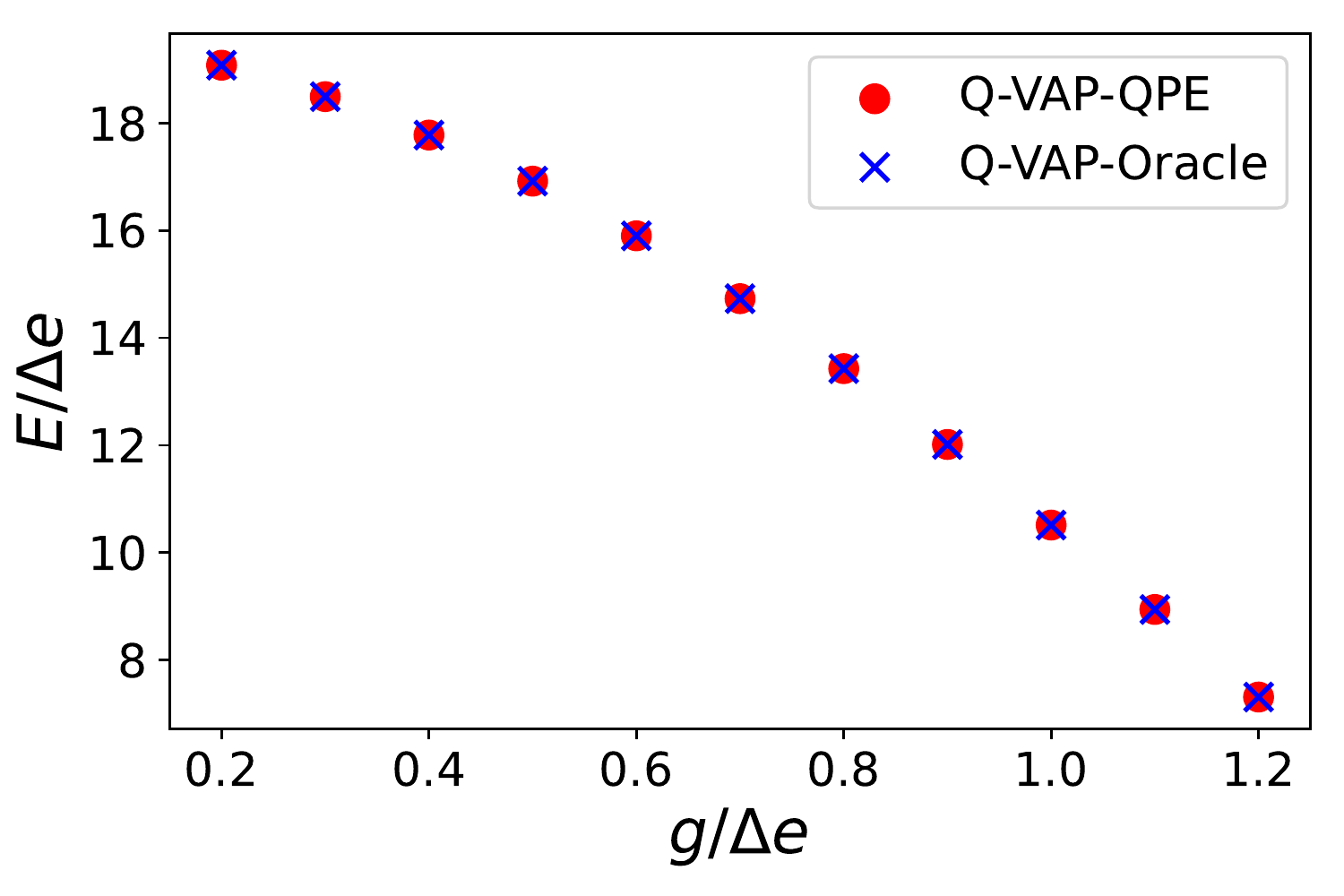}    
\end{center}
\caption{Illustration of the approximate ground state energy
obtained using the Q-VAP approach of Ref. \cite{Rui22} where the projection is made either 
by the indirect measurement technique as proposed in Ref. \cite{Lac20} (red filled circles) or by the oracle method of the present work (blue crosses). }
\label{fig:q_vap_pairing_8_qubits}
\end{figure}

We finally repeated the calculation performed in Ref. \cite{Rui22} where the Q-VAP approach was applied to a set of fermions
with equidistant level spacing $\varepsilon_i = i \Delta e$, interacting with each other through a pairing interaction whose strength was denoted by $g$.
In this reference, a BCS state parameterized with a set of parameters $\{ \theta_i\}_{i=1,8}$ was prepared on a quantum register. 
This state was then projected onto a given particle number using the method proposed in Ref. \cite{Lac20} based on measuring a set of ancillary qubits. 
The projected state was then optimized using quantum-classical hybrid calculations leading to the Q-VAP minimum of energy. 
The Q-VAP procedure was repeated here using the oracle method for projecting the particle number symmetry. 
The previous and new results are compared in Fig. \ref{fig:q_vap_pairing_8_qubits}. 
All details regarding the superfluid system encoding, state preparation, and
optimization technique are given in Ref. \cite{Rui22}. 
We see that the energies obtained with the two projection techniques perfectly match, 
with the advantage that the oracle-based approach requires only one ancillary qubit and that there are no rejected measurements.

\section{Conclusion}

In the present exploratory study, we discuss the possibility of using the concept of an oracle for the restoration of symmetries in many-body systems.
We illustrate using examples that the indirect measurement technique used in \cite{Rui22} to perform the Quantum Variation After projection can be replaced by the oracle projection method.
These examples put on evidence that the oracle also provides an accurate technique to perform projection with the advantage of not requiring the explicit construction of the projected state,
reducing the number of ancillary qubits and avoiding rejecting unwanted events.

Additionally, we show how oracles and projectors associated with symmetries can be constructed in practice on a quantum computer by first decomposing the projectors into a weighted sum of unitary operators.
Once such decomposition is achieved, a standard technique based on the LCU method can be envisaged for implementing projectors or oracles.
If we consider using the LCU technique, it is advisable to directly use the projector given that we would not have to intricate an entire operator with the ancilla qubit of the Hadamard test. 
On the other hand, if we can efficiently encode the oracle on the quantum circuit, its use would be preferable.

We finally mention that the construction of an oracle associated with projectors either by the LCU method or 
by the direct encoding of a diagonal operator remains a challenge for current noisy machines and that the method will only be applicable in the post-NISQ period.

\subsection*{Acknowledgments }
This project has received financial support from the CNRS through
the 80Prime program and is part of the QC2I project. We acknowledge
the use of IBM Q cloud as well as use of the Qiskit software package
\cite{Ibm21} for performing the quantum simulations.

\appendix 

\section{Technical aspects for the total spin projection}

\label{app:s2}

To implement the circuit associated with the spin projection, we first rewrite $\hat{S}^{2}$ given by Eq. (\ref{eq:s_2}) as:
\begin{eqnarray}
\hat{S}^{2} &=& \frac{3n_q}{4}+\frac{1}{2}\sum_{i<j,j=0}^{n_q-1}\left(X_{i}X_{j}+Y_{i}Y_{j}+Z_{i}Z_{j}\right).
\end{eqnarray}
The projector given in Eq (\ref{eq:reduced_projector_nb_particles}), written in the specific case where we project on the spin value $S'$ becomes:
\begin{align*}
 \hat{P}_{S'} & =  \sum_{k=0}^{n_q}\alpha_{k}e^{i\frac{\phi_{k}}{2}\sum_{i<j,j=0}^{n_q-1}\left(X_{i}X_{j}+Y_{i}Y_{j}\right)}e^{i\frac{\phi_{k}}{2}\sum_{i<j,j=0}^{n_q-1}\left(Z_{i}Z_{j}\right)} \nonumber \\
&  = \sum_{k=0}^{n_q} \alpha_{k}e^{i\frac{\phi_{k}}{2}\sum_{i<j,j=0}^{n_q-1}\left(X_{i}X_{j}+Y_{i}Y_{j}\right)}\prod_{i<j,j=0}^{^{n_q-1}}e^{i\frac{\phi_{k}}{2}Z_{i}Z_{j}} \nonumber 
\end{align*}
with $\alpha_{k}=\frac{e^{i\phi_{k}\left[\frac{3n_q}{4}-\xi_{S'}-\lambda_1\right]}}{M+1}$ and $\phi_k=\frac{2\pi k}{a\left(M+1\right)}$.
We recognize the $R_{ZZ}\left(\theta\right)=e^{-i\frac{\theta}{2}Z\otimes Z}$
gate; a decomposition of this gate in elemental gates is shown in
Fig \ref{fig:decomposition_rzz}. Since the terms $X_{i}X_{j}+Y_{i}Y_{j}$
do not commute; we use the Trotter-Suzuki approximation as:
\begin{equation}
e^{i\frac{\phi_{k}}{2}\sum_{i<j,j=0}^{n_q-1}\left(X_{i}X_{j}+Y_{i}Y_{j}\right)}\approx\left(\prod_{i<j,j=0}^{^{n_q-1}}e^{i\frac{\phi_{k}}{2n_{t}}\left(X_{i}X_{j}+Y_{i}Y_{j}\right)}\right)^{n_{t}}, \label{eq:trotter_xx_yy}
\end{equation}
where $n_{t}$ is the number of Trotter-Suzuki steps. A circuit to implement
an operator $e^{-i\frac{\theta}{4}\left(X\otimes X+Y\otimes Y\right)}$
is shown in Fig \ref{fig:decomposition_xx_yy}. $\theta$ should be
equal to $-\frac{2\phi_{k}}{n_{t}}$ in order to implement the gates
in Eq (\ref{eq:trotter_xx_yy}).

\begin{figure}[htbp]
\begin{centering}
\begin{tikzpicture}   
    \node[scale=1] { 
\begin{quantikz}  
\qw   & \gate[wires=2]{R_{zz}(\theta)} & \qw & \\
\qw   &                                & \qw
\end{quantikz} 
=\begin{quantikz}  
\qw   & \ctrl{1} & \qw                 &  \ctrl{1} & \qw \\
\qw   & \targ{}  & \gate{R_z(\theta)}  &  \targ{}  & \qw
\end{quantikz} 
}; 
\end{tikzpicture}\\
\par\end{centering}
\centering{}\caption{Decomposition of the $R_{ZZ}\left(\theta\right)=e^{-i\frac{\theta}{2}Z\otimes Z}$
gate. \label{fig:decomposition_rzz}}
\end{figure}
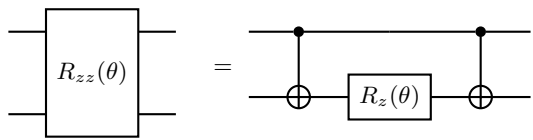

\begin{figure}[htbp]
\begin{centering}
\begin{tikzpicture}   
    \node[scale=1] { 
\begin{quantikz}  
\qw   & \targ{}   & \ctrl{1}            &  \targ{}   & \qw \\
\qw   & \ctrl{-1}  & \gate{R_x(\theta)}  &  \ctrl{-1}  & \qw
\end{quantikz} 
}; 
\end{tikzpicture}\\
\par\end{centering}
\centering{}\caption{Circuit to implement the operator $e^{-i\frac{\theta}{4}\left(X\otimes X+Y\otimes Y\right)}$.
\label{fig:decomposition_xx_yy}}
\end{figure}
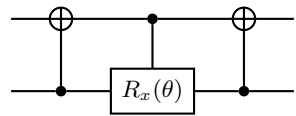

\bibliographystyle{unsrt}

\end{document}